\documentclass[10pt,aps,pra,twocolumn,superscriptaddress,floatfix,showpacs,longbibliography]{revtex4-1}
\usepackage{graphicx}
\usepackage{amsfonts}
\usepackage{amssymb}
\usepackage{amsmath}
\usepackage{txfonts}
\usepackage{lipsum}
\usepackage{color}
\usepackage{wasysym}
\usepackage{hyperref}
\usepackage{bbold}
\usepackage{etoolbox} % for additional refs as footnotes
\graphicspath{{figures/}}

\usepackage{mathtools}%enhanced math capabilities
\usepackage{multirow}%for the tables
\usepackage{hhline}

\DeclareMathOperator{\Tr}{Tr}

\newcommand{\qv}{\mathbf{q}}
\newcommand{\kv}{\mathbf{k}}

\usepackage{letltxmacro}
\LetLtxMacro{\oldsqrt}{\sqrt}
\renewcommand{\sqrt}[2][\mkern8mu]{\mkern-6mu\mathop{}\oldsqrt[#1]{#2}}

\begin{document}
\title{Resonant optical second harmonic generation in graphene-based heterostructures}

\author{M. Vandelli}
\affiliation{Radboud University, Institute for Molecules and Materials, 6525AJ Nijmegen, The Netherlands}
\affiliation{\mbox{Department of Physics, Informatics and Mathematics,
University of Modena and Reggio Emilia, 41125 Modena, Italy}}
%\affiliation{Max Planck Institute for the Structure and Dynamics of Matter, Center for Free Electron Laser Science, 22761 Hamburg, Germany}

\author{M. I. Katsnelson}
\affiliation{Radboud University, Institute for Molecules and Materials, 6525AJ Nijmegen, The Netherlands}
\affiliation{\mbox{Theoretical Physics and Applied Mathematics Department, Ural Federal University, Mira Street 19, 620002 Ekaterinburg, Russia}}

\author{E. A. Stepanov}
%\email{e.stepanov@science.ru.nl}
\affiliation{Radboud University, Institute for Molecules and Materials, 6525AJ Nijmegen, The Netherlands}
\affiliation{\mbox{Theoretical Physics and Applied Mathematics Department, Ural Federal University, Mira Street 19, 620002 Ekaterinburg, Russia}}

\begin{abstract}
An optical Second-Harmonic Generation (SHG) allows to probe various structural and symmetry-related properties of materials, since it is sensitive to the inversion symmetry breaking in the system. Here, we investigate the SHG response from a single layer of graphene disposed on an insulating hexagonal Boron Nitride (hBN) and Silicon Carbide (SiC) substrates. The considered systems are described by a non-interacting tight-binding model with a mass term, which describes a non-equivalence of two sublattices of graphene when the latter is placed on a substrate. The resulting SHG signal linearly depends on the degree of the inversion symmetry breaking (value of the mass term) and reveals several resonances associated with the band gap, van Hove singularity, and band width. The difficulty in distinguishing between SHG signals coming from the considered heterostrusture and environment (insulating substrate) can be avoided applying a homogeneous magnetic field. The latter creates Landau levels in the energy spectrum and leads to multiple resonances in the SHG spectrum. Position of these resonances explicitly depends on the value of the mass term. We show that at energies below the band-gap of the substrate the SHG signal from the massive graphene becomes resonant at physically relevant values of the applied magnetic field, while the SHG response from the environment stays off-resonant.
\end{abstract}

\maketitle

\section{Introduction}

The Second-Harmonic Generation (SHG) has become a very important tool to investigate different properties of materials. The sensitivity of the SHG to inversion symmetry and number of layers is an important aspect to perform experiments or to realize devices based on quasi-two-dimensional (2D) heterostructures~\cite{expTMDs}. The fact that the SHG is forbidden in materials where the inversion symmetry is preserved~\cite{boyd} can also be exploited for the investigation of layered systems composed from different materials. Recently, such heterostructures attracted a lot of attention from the physical community due to their unusual electronic properties~\cite{2dmat}. The state-of-the art method to study structural properties of these systems is the scanning tunneling microscopy (STM). However, the direct STM measurements can probe only the surface states and are not sensitive to structural changes in multilayered heterostructures. On the other hand a much simpler experiment on the SHG can indirectly capture the differences between various combinations of layers even if it occurs not at the surface.

Unfortunately, an application of this technique to a simplest and most extensively studied 2D material, i.e. a monolayer graphene, turns out to be inefficient. Indeed, a pristine graphene exhibits inversion symmetry, which prevents any SHG.
The SHG signal in graphene can be observed either by inducing an asymmetry between two sublattices of graphene placing it on top of a band insulator, or considering the fact that a photon momentum $\qv$ of the applied light already works as a source of asymmetry. It has been shown that the response caused by the photon momentum is weak, since it is proportional to the momentum itself~\cite{Glazov2011}. The problem of the SHG in the case of the hexagonal lattice with the broken inversion symmetry has also been considered previously~\cite{ultrastrong, marg}. It is worth mentioning another physical effect, namely the valley polarization, that allows the SHG in graphene~\cite{valleypol}.
Although this mechanism could be very useful in the context of \emph{valleytronics}, addressing the valley polarization experimentally is still a matter of a research.

In this work we investigate the SHG from the graphene disposed on the insulating hexagonal substrates with different band-gaps as a particular example of quasi-2D heterostructures mentioned above. For this aim we perform calculations using the diagrammatic technique based on the full dispersion of the non-interacting tight-binding model with the mass term. The reason for a yet another theoretical study of the SHG in graphene with the broken inversion symmetry is the fact that previous studies on this subject report features that can hardly be explained within physical intuition. For instance, this concerns a stronger SHG response for smaller mass term~\cite{marg}, which is very surprising, since the mass is a consequence of the inversion symmetry breaking. Furthermore, an overwhelming majority of theoretical studies of the SHG in graphene-based heterostructures are focused on frequencies of the incident light around the band-gap, which is far from the experimentally accessible regime, where the energy of the incoming photons is usually around 1.5\,eV for red light sources. For this reason, we obtain the full SHG spectrum that is needed for a description of the actual experimental data. This allows us to reveal additional resonances in the optical spectrum that correspond to the van Hove singularity and band-width, which cannot be captured by a simplified Dirac model. The comparison between the SHG response obtained for the full tight-binding spectrum and the one for the approximated Dirac picture allows us to define the limits of applicability of this approximation.

Another experimentally relevant problem that stays undiscussed in all previous works is the difficulty to distinguish between SHG signals from the graphene flake placed on the insulating substrate and the rest of the insulating sample. Indeed, we find that, contrary to the result of the Ref.~\onlinecite{marg}, the SHG response is proportional to the band-gap, which in the case of graphene disposed on the substrate is small compared to the band-gap of the clean substrate. Therefore, the SHG signal from graphene can hardly be seen on top of the large SHG signal from the band insulator. We show that this problem can be resolved including the homogeneous magnetic field in the system. The presence of the magnetic field results in the formation of Landau levels in the energy spectrum. Since these levels are sharp, we expect intense resonances associated with transitions between Landau levels in the SHG spectrum. Therefore, the presence of the magnetic field introduces a natural amplification of the SHG that can be tuned adjusting the value of the external magnetic field at a fixed laser frequency. The small value of the mass term of the graphene-based hetrostructure allows to find resonances on Landau levels already at energies below the band-gap of the insulating substrate, while the SHG signal of the environment stays off-resonant.
%This fact allows to detect the presence of the graphene flake on the insulating sample.
An experimental evidence of the applicability of this technique can be found in Ref.~\onlinecite{SHGGaAs}.

\section{SHG response from massive graphene}

\begin{figure}[t!]
\centering
\includegraphics[width=0.9\linewidth]{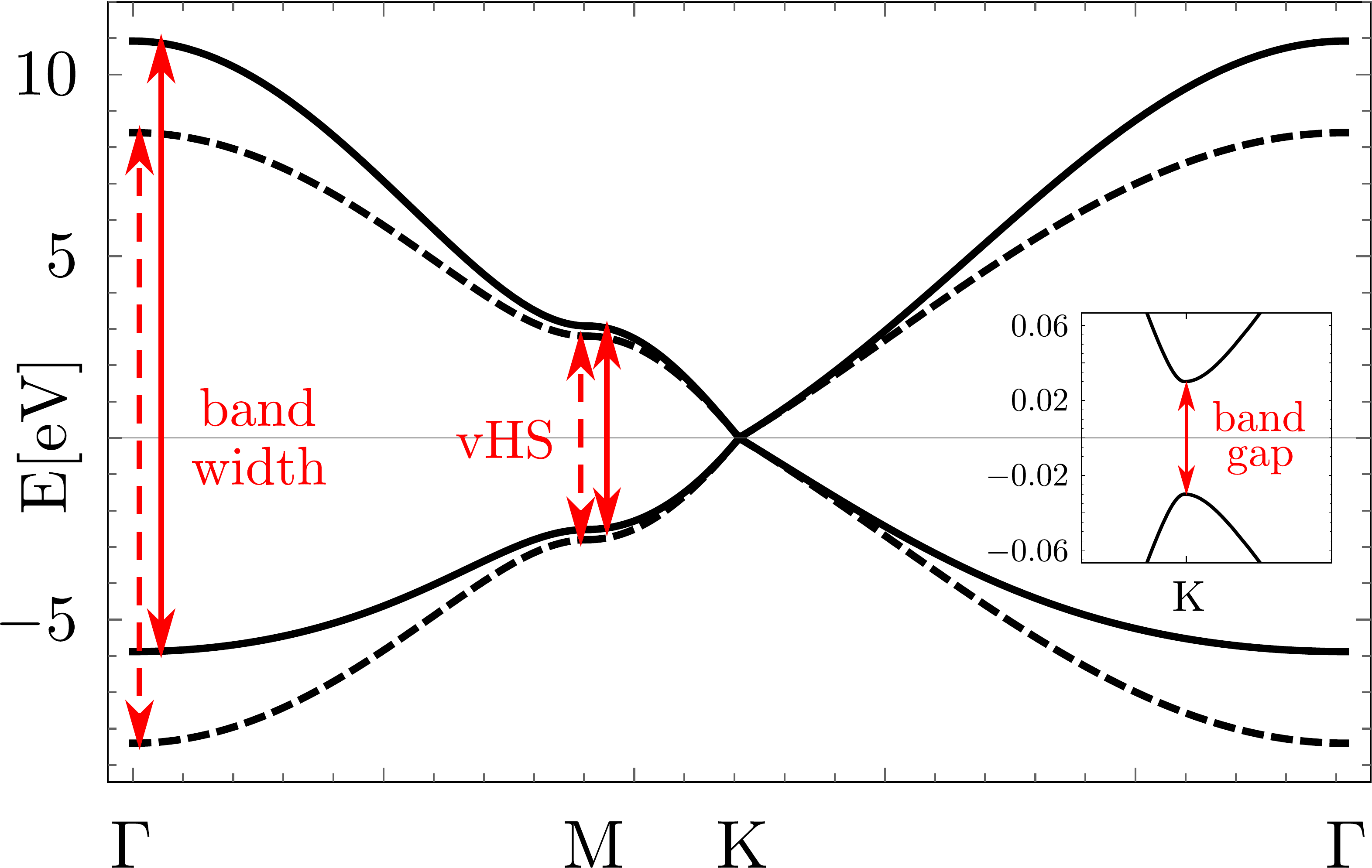}~
\caption{\label{fig:dispersion} Dispersion relation of graphene with (solid line) and without (dashed line) account for the next-nearest-neighbor hopping process $t'$. Red arrows show optical resonances at the bandwidth ($\Gamma$ point), van Hove singularity (M point), and band-gap (K point).}
\vspace{-10pt}
\end{figure}

Here, we study optical second harmonic generation from graphene-based heterostructures using the following tight-binding model that describes a behavior of non-interacting electrons on a hexagonal lattice. The corresponding Hamiltonian matrix written in the sublattice space reads
\begin{align}
\label{hexham}
\hat{H}_{\bf k} =
\left(\begin{matrix}
   f_{\bf k}+m     & S_{\kv} \\
    S^{*}_{\kv}       & f_{\kv}-m \\
\end{matrix}\right).
\end{align}
Here, the off-diagonal term $S_{\kv}$ is a Fourier transform of the nearest-neighbour hopping process $t$, and the diagonal one $f_{\kv}$ describes the next-nearest-neighbor hoping of electrons $t'$ (see Appendix~\ref{app:diagrams}). Here we also introduce the mass term $m$ that explicitly breaks the inversion symmetry of the system.
The mass term describes the sublattice imbalance, which illustrates the situation when graphene is placed on top of a band insulator,
such as the hexagonal Boron Nitride (hBN)~\cite{woods2014commensurate, PhysRevB.87.245408, hBNgap, hBNrelax}, or Silicon Carbide (SiC)~\cite{zhou}, as discussed above. This model is found to be relevant for other materials, such as the monolayer MoS$_2$~\cite{PhysRevB.87.245421} and germanene under the effect of an electric field~\cite{0953-8984-27-44-443002}. Note that for the case of graphene on hBN the appearance of the gap corresponds to the commensurate phase at small enough misorientation angle between their lattices, whereas for the incommensurate phase the average gap is supposed to be zero~\cite{woods2014commensurate, titov2014}. It is the average gap that matters in SHG experiments, where the typical laser spot size is much larger than the interatomic distance. In the following when discussing graphene placed on the insulating substrate we will keep in mind only the commensurate case.
The dispersion relation
$E_{\bf k}^{\pm} = f^{\phantom{|}}_{\bf k} \pm \sqrt{|S^{\phantom{|}}_{\bf k}|^2+m^2}$
for the case of graphene on top of hBN with $t=-2.8$ eV, $t'=-0.1t$~\cite{PhysRevB.88.165427}, and $m=30$ meV is shown in Fig.~\ref{fig:dispersion}.

Following the procedure adopted from Ref.~\onlinecite{valleypol}, the effect of the applied probe light is accounted via the Peierls substitution introducing a vector potential ${\bf A}$ that represents an external radiation
\begin{align}
\hat{H}_{ij}[A] &=  \hat{H}_{ij}\exp\left(-i\frac{e}{c}\int_{{\bf R}_i}^{{\bf R}_j} {\bf A}({\bf r},t)\cdot d{\bf r}\right).
\label{tightbHamRad}
\end{align}
Here, $e$ is the modulus of the electronic charge, $c$ is the speed of light, and $\hat{H}_{ij}$ is the lattice $\{i, j\}$ space representation of the Hamiltonian matrix~\eqref{hexham}.

The SHG response function we aim to obtain in the current work can be derived using the Feynman diagram technique. For this reason, we stick to the path integral formalism with the corresponding action for our problem
\begin{align}
{\cal S}[A] = -\Tr\sum_{\kv\nu} \hat{c}^{*}_{\kv\nu}\left(\mathbb{1}\left(i\nu + \mu\right) - \hat{H}_{\kv}[A] \right) \hat{c}^{\phantom{*}}_{\kv\nu}.
\label{eq:action}
\end{align}
Here, $\hat{c}^{(*)}_{\kv}$ is a two-dimensional spinor of Grassman variables that describe creation (annihilation) of an electron, and $\mathbb{1}$ is the $2\times2$ identity matrix
%and  $\boldsymbol{\sigma}=(\sigma_x,\sigma_y,\sigma_z)$ is the vector of Pauli Matrices
in the sublattice space; the trace is taken over the same space, $\beta$ is the inverse temperature and $\sum_{\kv,\nu}$ stands for the infinite summation over the fermionic Matsubara frequencies $\nu_{n}=(2n+1)\pi/\beta$ and momentum integration over the Brillouin Zone ($BZ$).
%We also define $\boldsymbol{\xi}_{\bf k} = (\mathrm{Re}\, S_\kv,\mathrm{Im}\, S_\kv,m)$.
The chemical potential $\mu=-3t'$ corresponds to the half-filled case (neutrality point). 
%\begin{align}
%\sum_{\kv,\nu} = \frac{1}{\beta}\sum_{n=-\infty}^{+\infty}\,\int\limits_{BZ} %\frac{d^2k}{(2\pi)^2}.
%\end{align}

The electric current density can be defined as the response of the system on the applied vector potential
$
j_{\alpha\omega}[A] = \delta{\cal F}[A]/\delta A_{\alpha\omega},
$
where ${\cal F}[A]=\ln{\cal Z}[A]$ and ${\cal Z}[A] = \int D[c^{*},c] \exp\left\{-\beta{\cal S}[A]\right\}$
is the is the generating functional, that is, the partition function of the problem written in terms of the action~\eqref{eq:action}.
Expanding the electric current up to the second order with respect to the vector potential, one gets the usual relation
\begin{align}
j_{\alpha\omega}[A] - j_{\alpha\omega}[0] &= \sum_{\beta,\omega'}\left.\frac{\delta j_{\alpha\omega}[A]}{\delta A_{\beta\omega'}}\right|_{A=0} A_{\beta\omega'} \\
&+ \frac{1}{2}
\sum_{\beta\gamma,\omega'\omega''}\left.\frac{\delta^2 j_{\alpha\omega}[A]}{\delta A_{\beta\omega'}\delta A_{\gamma\omega''}}\right|_{A=0} A_{\beta\omega'}A_{\gamma\omega''}. \notag
\end{align}
The coefficient in front of the liner term is the usual optical conductivity of the system. The coefficient in front of the square of the vector potential describes the second harmonic generation. The latter can also be expressed via the three-particle correlation function $\Pi_{\alpha\beta\gamma}^{\omega\omega'\omega''}$ as
\begin{align}
\left.\frac{\delta^2 j_{\alpha\omega}[A]}{\delta A_{\beta\omega'}\delta A_{\gamma\omega''}}\right|_{A=0} = \left.\frac{\delta^3 {\cal F}[A]}{\delta A_{\alpha\omega}A_{\beta\omega'}\delta A_{\gamma\omega''}}\right|_{A=0} = 2 e^3\Pi_{\alpha\beta\gamma}^{\omega\omega'\omega''}.
\label{eq:diagrams}
\end{align}

\begin{figure}[t!]
\includegraphics[width=1\linewidth]{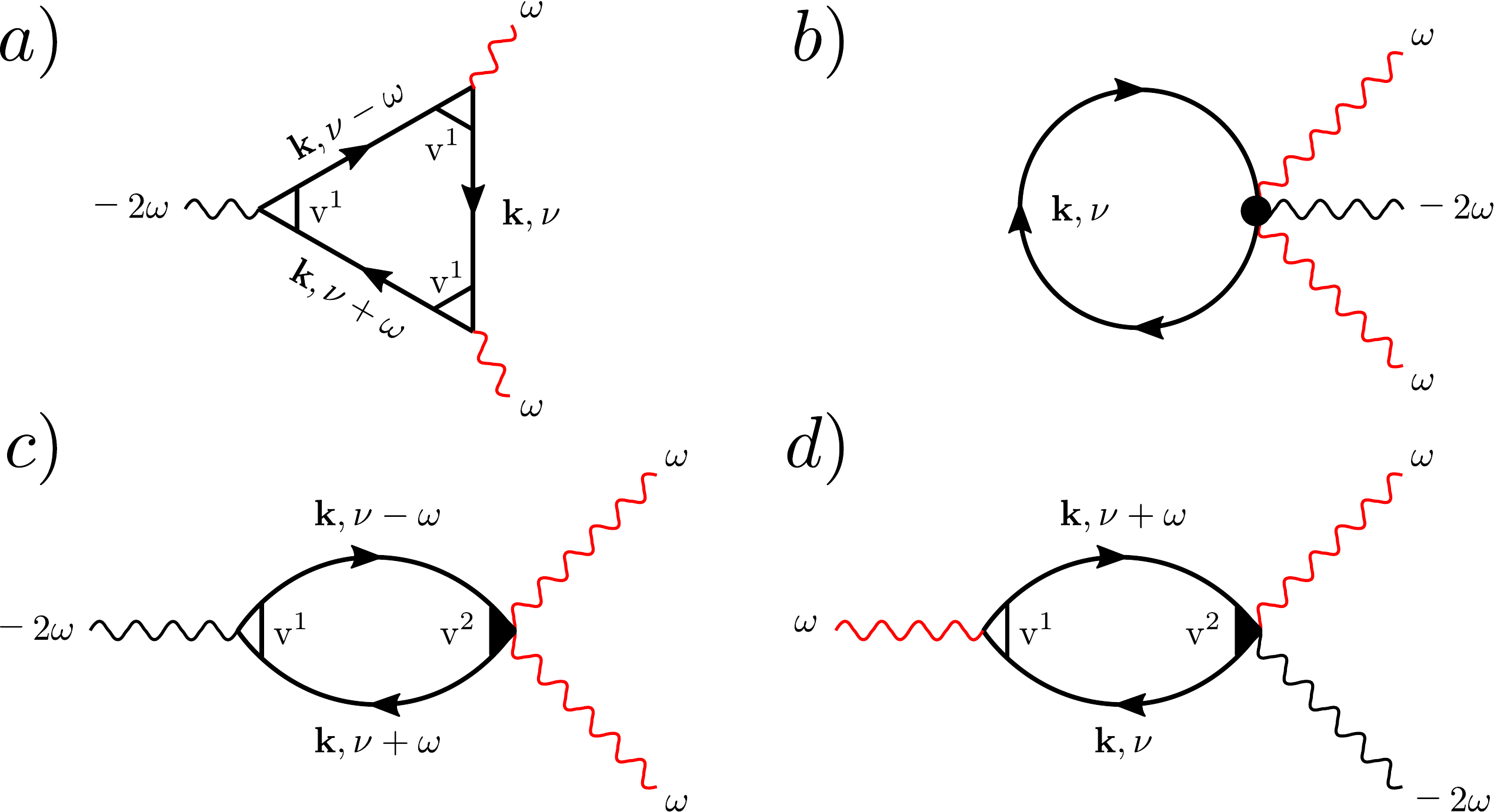}
\caption{The triangular a), frequency independent b), and non-linear bubble diagrams c)-d) involved in the SHG process.\label{fig:diag}}
\end{figure}

Diagrammatic expressions for this correlation function are shown in Fig.~\ref{fig:diag}. The diagram ``$b$'' will be ignored in the following, because it represents a constant energy shift. Explicit expressions for the triangular $\Pi^{(3)}(\omega)$ (``$a$'') and nonlinear bubble $\Pi^{(2)}(\omega)$ (``$c$'' and ``$d$'') diagrams are following
\begin{align}
\label{diag}
\Pi^{(2)}_{\alpha\beta\delta}(\omega) &= \Tr\sum\limits_{\kv, \nu} \hat{\rm v}^{(2)}_{\alpha\beta} \hat{G}(\textbf{k},\nu-\omega){\rm v}^{(1)}_\delta \hat{G}(\textbf{k},\nu+\omega) \\
&+ 2\Tr\sum_{{\kv,\nu}} \hat{\rm v}^{(2)}_{\alpha\beta}  \; \hat{G}\left({\kv,\nu+\omega}\right) \, \hat{\rm v}^{(1)}_{\gamma}  \; \hat{G}\left({\kv,\nu}\right), \notag\\
\Pi^{(3)}_{\alpha\beta\gamma}(\omega) &= \Tr\sum\limits_{\kv,\nu}  \hat{\rm v}^{(1)}_\alpha \hat{G}(\kv,\nu + \omega) \hat{\rm v}^{(1)}_\beta \hat{G}(\kv, \nu) \hat{\rm v}^{(1)}_\gamma \hat{G}(\kv,\nu -\omega), \notag
\end{align}
where $\hat{G}({\bf k}, \nu) = \left[\mathbb{1}(i\nu + \mu)- \hat{H}_{\bf k}\right]^{-1} $ is the Green's function of our problem, and velocity operators can be defined in the same way as in Ref.~\onlinecite{valleypol}
\begin{align}
\hat{\rm v}^{(1)}_{\alpha} =  \left.\frac{1}{e}\frac{\delta \hat{H}_{\kv}[A]}{\delta A_{\alpha}}\right\rvert_{A=0},~~~~
\hat{\rm v}^{(2)}_{\alpha\beta} =  \left.\frac{1}{e^{2}}\frac{\delta^{2} \hat{H}_{\kv}[A]}{\delta A_{\alpha} \delta A_{\beta}}\right\rvert_{A=0}.
\end{align}

As can be seen from Eq.~\ref{diag}, expressions for diagrams ``$c$'' and ``$d$'' are connected by the following simple relation $\Pi_{\alpha\beta\gamma}^{(2)\,c}(\omega)=\Pi_{\alpha\beta\gamma}^{(2)\,d}(2\omega)$.
The coefficient $2$ is not included in the definition of $\Pi_{\alpha\beta\gamma}^{(2)\,d}(2\omega)$.
Then, the total result for the nonlinear bubble can be written as $\Pi_{\alpha\beta\gamma}^{(2)}(\omega) = \Pi_{\alpha\beta\gamma}^{(2)\,d}(2\omega) + 2\Pi_{\alpha\beta\gamma}^{(2)\,d}(\omega)$, which explicitly connects the behavior of the SHG spectrum at double- and single frequencies of the applied light. It is also worth mentioning that the contribution of the $\Pi_{\alpha\beta\gamma}^{(2)\,d}(\omega)$ diagram is missing in~\cite{valleypol}.

The real frequency dependence of correlation functions can be obtained performing an analytic continuation
\begin{align}
\bar{\Pi}_{\alpha\beta\gamma}(\omega) = \lim\limits_{\epsilon \rightarrow 0} \left[ \Pi_{\alpha\beta\gamma}(-i\omega + \epsilon) - \Pi_{\alpha\beta\gamma}(0)\right].
\end{align}
In practical calculations $\epsilon$ is taken to be small, but finite.

In the following we take into account that the experimentally measurable quantity for the SHG is the conversion efficiency. It can be defined in the same way as in Ref.~\onlinecite{boyd} and is proportional to the ratio
\begin{align}
\eta(\omega) = \bar{\Pi}(\omega)/\omega.
\end{align}

The explicit evaluation of the introduced diagrams is shown in the Appendix~\ref{app:tensor}. We find that the contribution of the triangular diagram $\Pi^{(3)}(\omega)$ is zero even when a non-zero next-nearest-neighbor hopping $t'$ and chemical potential away from the half-filling are considered.
The reason is that the integral over momentum $\kv$ in equation~\eqref{diag} for $\Pi^{(3)}(\omega)$ averages to zero in the whole Brillouin zone. This is essentially due to the fact that two valleys $K$ and $K'$ of graphene contribute to the integral with opposite signs and hence compensate each other.
This result is a generalization of the situation considered in~\cite{ultrastrong} for the case of a low-energy Hamiltonian for MoS$_2$ material, where the contribution of the triangular diagram is canceled by symmetry with respect to the inversion of $k_y$.
A non-zero result for the triangular diagram can be obtained introducing a valley polarization that generates an imbalance between the two valleys, as discussed in Ref.~\onlinecite{valleypol}.

Contrary to the triangular diagram, the contribution from the nonlinear bubble is nonzero, and the $\Pi^{(2)}_{\alpha\beta\gamma}(\omega)$ tensor reveals the reduced symmetry $C_{3}$ instead of $C_{6}$ with respect to rotation (see Appendix~\ref{app:tensor}). Thus, we find that the contribution $\Pi^{(2)}_{xxx}(\omega)=0$, whereas the result for $\Pi^{(2)}_{yyy}(\omega)$ is nonzero.
It can be shown that the only non-zero components of the tensor are $\Pi^{(2)}(\omega) = \Pi_{yyy}^{(2)}(\omega)=-\Pi_{xxy}^{(2)}(\omega)=-\Pi_{yxx}^{(2)}(\omega)=-\Pi_{xyx}^{(2)}(\omega)$.

Remarkably, the account for the next-nearest-neighbor hopping process $t'$ also does not change the result for the nonlinear bubble diagram. This can be explained looking at the dispersion relation in Fig.~\ref{fig:dispersion}. The inclusion of $t'$ equally shifts the upper and the lower band at given momentum $\kv$, which does not change the energy difference between them. Since we consider only direct excitations at zero momentum, the SHG spectrum depends on energy difference between two bands and hence does not change with the inclusion of the next-nearest-neighbor hopping.

\begin{figure}[t!]
\includegraphics[width=1\linewidth]{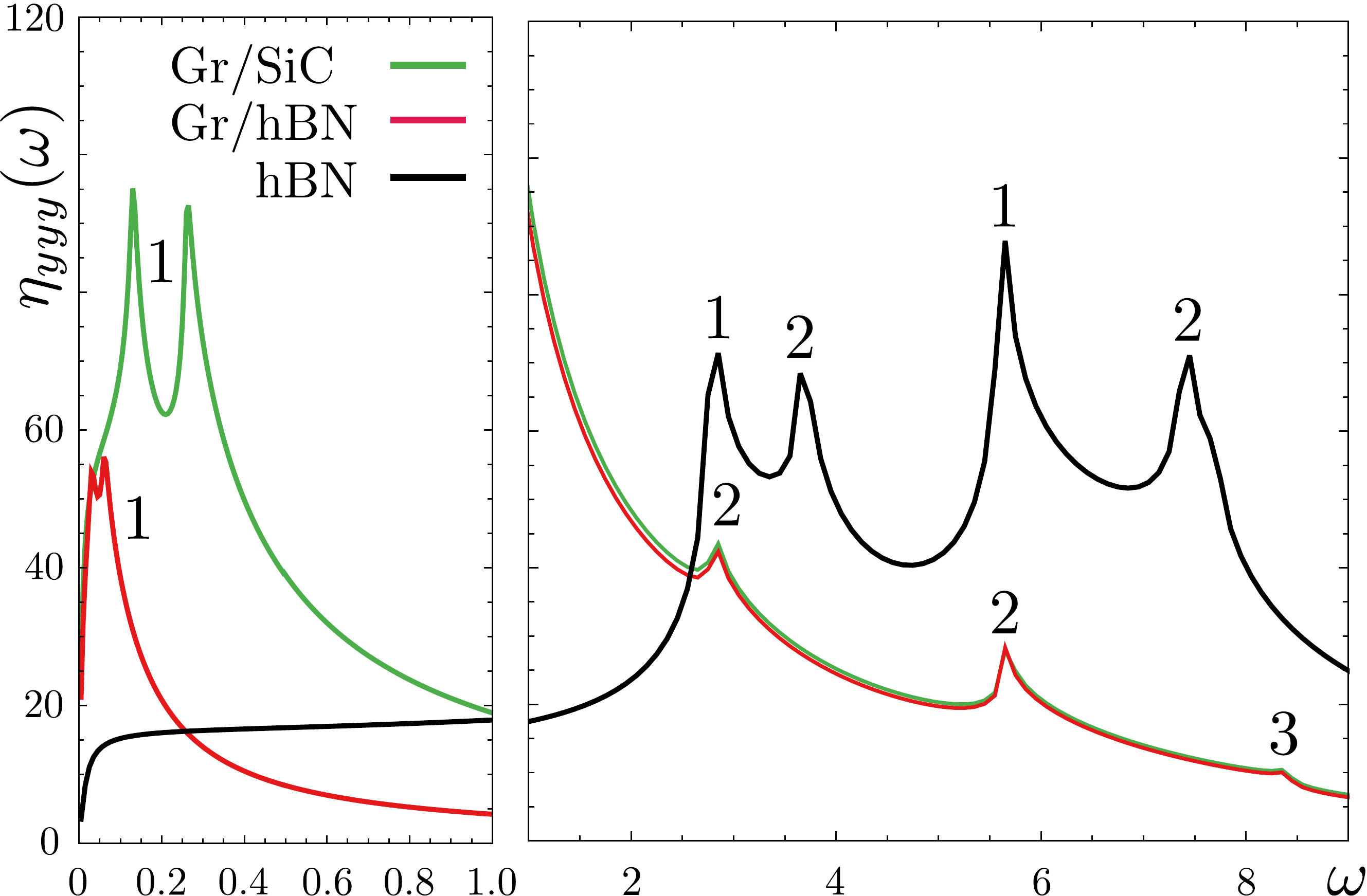}~
\caption{The absolute value of $\eta_{yyy}(\omega)$ for hBN (black line), Gr/SiC (green line) and Gr/hBN (red line) at low (left) and high (right) frequency $\omega$. The data for Gr/SiC on the right panel is multiplied by a factor of 5 and data for Gr/hBN is multiplied by $5\times(m_{\rm Gr/SiC} / m_{\rm Gr/hBN})$. The data on the left panel is shown without the multiplication. Labels ``1'', ``2'', and ``3'' depict resonances on the band-gap, van Hove singularity, and the band width, respectively.}
\label{fig:PiSHG}
\end{figure}

Fig.~\ref{fig:PiSHG} shows the absolute value of the conversion efficiency $\eta_{yyy}(\omega)$ for different values of the mass term (half of the band-gap). Here, the results for the hexagonal Boron Nitride (hBN, $t=-2.4$ eV, $m = 2.78$ eV), graphene on hBN (Gr/hBN, $t=-2.8$ eV, $m=0.03$ eV), and graphene on a SiC substrate (Gr/SiC, $t=-2.8$ eV, $m=0.13$ eV)~\cite{hBNgap, marg} are compared. The data on the right panel for the Gr/SiC is multiplied by a factor of 5, and the one for the Gr/hBN is multiplied by $5\times(m_{\rm Gr/SiC} \div m_{\rm Gr/hBN})$ for clarity. The real and imaginary parts of the conversion efficiency $\eta_{yyy}(\omega)$ are shown in Appendix~\ref{app:tensor}.

The SHG is a virtual process that is allowed even if the frequency of the applied light is smaller than the half of the band-gap. The nonzero increasing value of the conversion efficiency $\eta(\omega)$ at small $\omega<m$ frequencies confirms this statement. This consideration appears to be in agreement with the result of~\cite{ultrastrong}, and also reproduces the trend observed in~\cite{PhysRevB.91.205405}. Increasing the frequency of the applied light, the SHG spectrum reveals the first resonance at energies when excited electrons reach the band-gap. This resonance appears as a pair of peaks at frequencies $\omega=m$ and $\omega=2m$ in agreement with the frequency dependence of the nonlinear bubble diagram $\Pi^{(2)}(\omega)$ discussed above. These peaks are labeled as ``1'' in Fig.~\ref{fig:PiSHG}. The double resonance on the band-gap was reported previously in~\cite{marg}, but is missing, for example, in~\cite{PhysRevB.91.205405}.

The use of the full tight-binding dispersion allows us to capture additional resonances in the SHG spectrum labelled as ``2'' and ``3'' in Fig.~\ref{fig:PiSHG}. The peak ``2'' corresponds to the van Hove singularity and appears in the nonlinear optical spectrum at the frequency $\omega=2.8$ eV (Gr/hBN and Gr/SiC) and $\omega=3.67$ eV (hBN) with its replica at $2\omega$. The resonant peak ``3'' at the highest energy comes from the bandwidth of the system. Note that the use of the full tight-binding dispersion is crucial for a description of these additional optical resonances depicted in Fig.~\ref{fig:dispersion} via red arrows, since the low energy expansion (Dirac picture) does not provide the corresponding features of the energy spectrum.

In addition, we observe that the off-resonant SHG response function linearly depends on the value of the mass term. Indeed, the conversion efficiency $\eta(\omega)$ for the Gr/SiC shown in Fig.~\ref{fig:PiSHG} (right) is almost indistinguishable from the one of the Gr/hBN multiplied by the factor $(m_{\rm Gr/SiC}/ m_{\rm Gr/hBN})$.
The fact that a smaller mass term leads to a smaller value of the conversion efficiency can also be seen comparing the off-resonant behavior of $\eta(\omega)$ for the hBN with the one for graphene-based heterostructures. For instance, the SHG response from the hBN for the applied red light ($\omega\simeq1.5$ eV) is five times larger than the one from the Gr/hBN. However, the direct comparison of these signals is complicated by a predominant resonant behavior of the conversion efficiency of hBN at large frequencies. From the physical point of view, this result can be explained as follows. The mass term is exactly the factor that breaks the inversion symmetry and hence is responsible for the SHG. Since the latter is identically zero in systems with the unbroken inversion symmetry, the larger SHG signal is expected when the symmetry breaking is more severe. 

It is worth mentioning that our result is in a disagreement with the one of the Ref.~\onlinecite{marg}, where the authors report a larger SHG signal for the material with a smaller mass term.

\section{Influence of magnetic field on the SHG}

Now, let us discuss the effect of a homogeneous magnetic field on the SHG.
The inclusion of the magnetic field in a general tight-binding model can be done only numerically, which is less intuitive and is much more complicated than an analytical solution of the problem. However, this issue can be resolved in the framework of the Dirac model. The latter can be obtained performing the low-energy expansion of the Hamiltonian matrix~\eqref{hexham} in the vicinity of $K$ and $K'$ points of the hexagonal Brillouin Zone of graphene~\cite{kats}
\begin{align}
\hat{H}_{\rm D} = v\left[\tau k_x \hat{\sigma}_x + k_y\hat{\sigma}_y\right] + m\hat{\sigma}_{z}.
\label{eq:DiracH}
\end{align}
Here, $\tau = \pm1$ is the valley index, and $v = 3at/2$ is the electron speed at conical points $K$ and $K'$.
%In absence of the mass term $m$ this leads to massless Dirac electrons in pristine Graphene~\cite{massless}.
%The finite mass term $m$ opens the gap $E_g = 2m$ in the energy spectrum, and the electrons behave like massive Dirac particles.

The SHG in the case of Dirac electrons is forbidden by Furry's theorem~\cite{peskin}. This is represented by the fact that the triangular diagram $\Pi^{(3)}(\omega)$ is identically zero. The nonlinear bubble diagram is absent in the Dirac approximation, since the corresponding low-energy Hamiltonian does not contain any second order term in momentum $\kv$ that is responsible for the existence of the the non-linear vertex ${\rm v}^{(2)}$. Moreover, the SHG response calculated on the basis of the low-energy Hamiltonian does not reproduce all features of the SHG spectrum shown in Fig.~\ref{fig:PiSHG}. Therefore, we need to go beyond the Dirac approximation in order to obtain an experimentally relevant result for the SHG. Here, we can benefit from the fact that the frequency of the red light, which is commonly used in SHG experiments, is smaller than the van Hove singularity. For this reason, the inclusion of already the first order correction in momentum $\kv$ to the low energy Hamiltonian~\eqref{eq:DiracH} will be sufficient to describe the SHG in graphene-based heterostructures, although the resulting model will not hold for energies around the van Hove singularities. We indicate the correction to the Diarc Hamiltonian as $H_{\text{TW}}$, which is the so called trigonal warping term
\begin{align}
\hat{H}_{\text{TW}} = \lambda \left[2\tau k_x k_y \hat{\sigma}_y  - (k_x^2 -k_y^2)\hat{\sigma}_x\right],
\end{align}
where $\lambda=3a^2t/8$ is the trigonal warping parameter~\cite{kats}.
The crucial role of the trigonal warping for the SHG in graphene at low energies was pointed out in previous works~\cite{golub, marg}, although the result for the SHG is obtained there for the regime of very low frequencies, which is not accessible experimentally. The Dirac approximation with the trigonal warping term was also considered in Ref.~\cite{ultrastrong} for the case of the SHG in MoS$_2$. From the diagrammatic point of view, the role of the trigonal warping is to introduce a non-linear vertex ${\rm v}^{(2)}$ in the theory, which is responsible for the existence of the non-linear bubble diagram $\Pi^{(2)}(\omega)$.
%Since the triangular diagram $\Pi^{(3)}(\omega)$ is found to be zero, the nonlinear bubble represents, in fact, the only contribution to the SHG.
In the following, we make an additional approximation expanding the $\Pi^{(2)}(\omega)$ up to the first order in the trigonal warping parameter $\lambda$. Then, the contribution of the trigonal warping remains only in the vertex function ${\rm v}^{2}$, while the Green's function stays the same as for massive Dirac electrons. This will allow to account for the effect of Landau levels in the Green's function analytically without any approximations.

\begin{figure}[t!]
\includegraphics[width=1\linewidth]{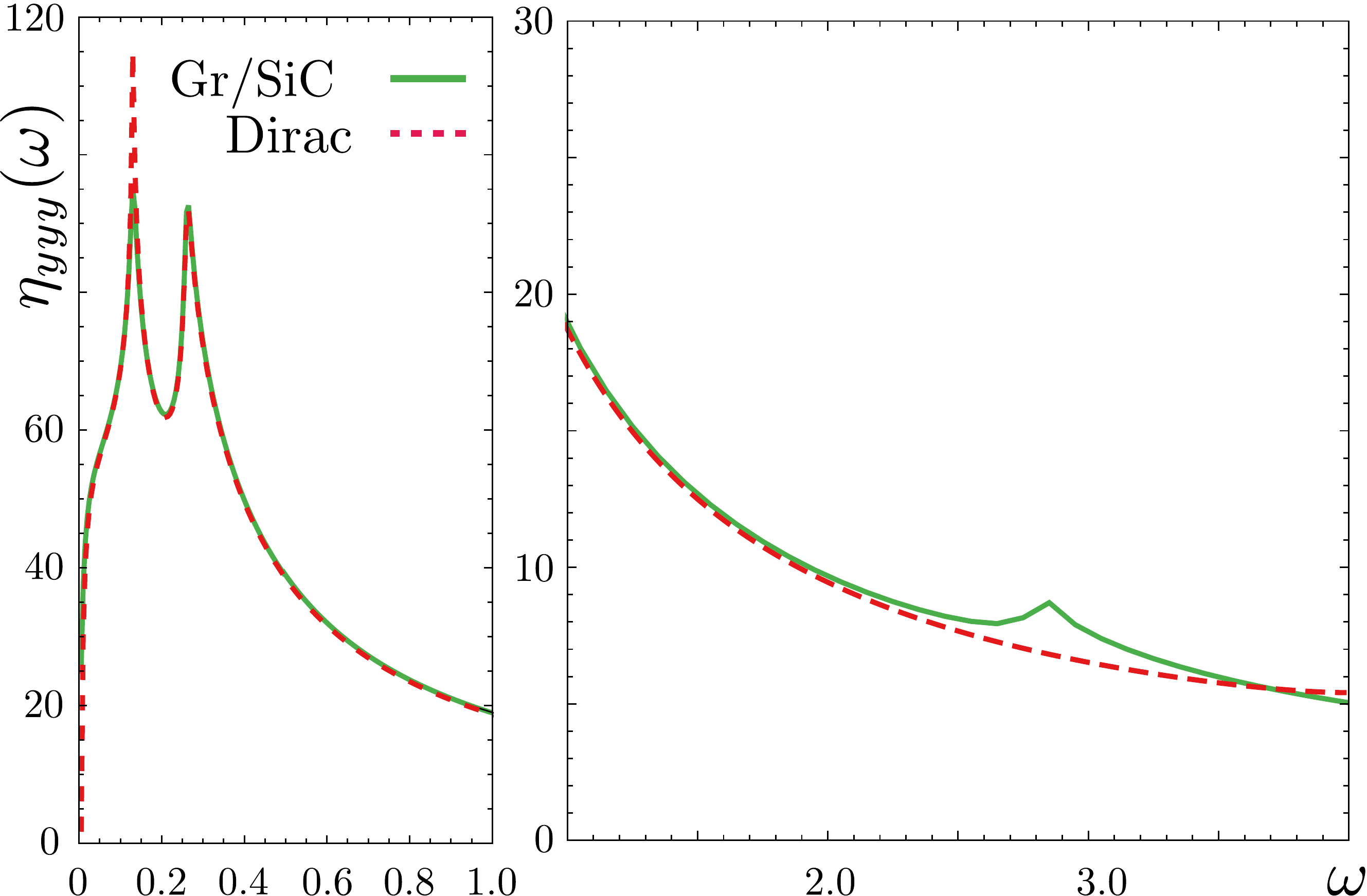}~
\caption{Comparison between absolute values of $\eta_{yyy}(\omega)$ for the full dispersion (solid line) and massive Dirac dispersion with trigonal warping (dashed line) for the case of Gr/SiC at zero value of the magnetic field.}
\label{fig:comparisonDir}
\end{figure}

In order to estimate limits of applicability of the derived approximation, let us consider the correction to the the dispersion relation due to the trigonal warping. At small values of the mass $m$, the contribution to the energy from Dirac dispersion is approximatelly equal to $\frac{3t}{2}k$ and the contribution from the trigonal warping term is $\frac{3t}{8}k^2$. The latter can be considered as a small correction for $k$ up to $0.4$, which corresponds to the energy of about $1.7$\,eV. Fig.~\ref{fig:comparisonDir} shows that this estimation is rather conservative, and the SHG response function of the approximate model is in a good agreement for the one obtained using the full tight-binding spectrum up to energies of about $2$\,eV. As expected, the breakdown of the approximation is associated with the presence of the resonance on the van Hove singularity in the SHG spectrum, which can not be reproduced without the full tight-binding dispersion.

After all, the conversion efficiency $\eta(\omega)$ at zero magnetic field can be recast in a very simple form (see Appendix~\ref{app:Dirac})
\begin{align}
\label{diracSHG}
\eta(\omega) = 12i m\lambda v \sum\limits_{\kv} \frac{\tanh\left(\frac{\beta \varepsilon_\kv}{2}\right)}{\varepsilon_\kv  }\left[\frac{1}{\omega^2-\varepsilon_\kv^2}+\frac{4}{\omega^2-4\varepsilon_\kv^2}\right],
\end{align}
where $\varepsilon_{\kv}=\sqrt{v^2 \kv^2+m^2}$ is the massive Dirac dispersion of electrons in graphene.
Remarkably, Eq.~\ref{diracSHG} shows that the off-resonant value of $\eta(\omega)$ linearly dependents on the
%trigonal parameter $\lambda$ and
mass term $m$, as expected from above discussions.
%It can also be seen that the temperature smears the peak corresponding to the band-gap by a factor of $\tanh \left(\frac{\beta\varepsilon_{\kv}}{2}\right)$. This leads to a suppression of the band-gap resonance for band-gaps \ES{of the order of the $1/\beta$, as for example in Gr/hBN.}

The homogeneous magnetic field directed perpendicular to the surface of the sample can be introduced in the system via the Peierls substitution ${\kv} \to \kv + e{\bf A}_B$  with the following vector potential ${\bf A}_B=\frac{B}{2}(-y,x,0)$. With this substitution, the energy spectrum of the problem changes dramatically from $\varepsilon_{\kv}$ to a discrete set of Landau levels described by the following expression $\varepsilon_n = \sqrt{m^2 + 2|eBv^2n|}$ with $n\in\mathbb{Z}$, (see~\cite{kats} and~\cite{Mir} for the case of zero mass term). %Noticeably this quantization of electronic levels in a magnetic field corresponds to a confinement to $d-2$ dimensions, as pointed out in Ref. \cite{Mir}.

\begin{figure}[t!]
\includegraphics[width=0.9\linewidth]{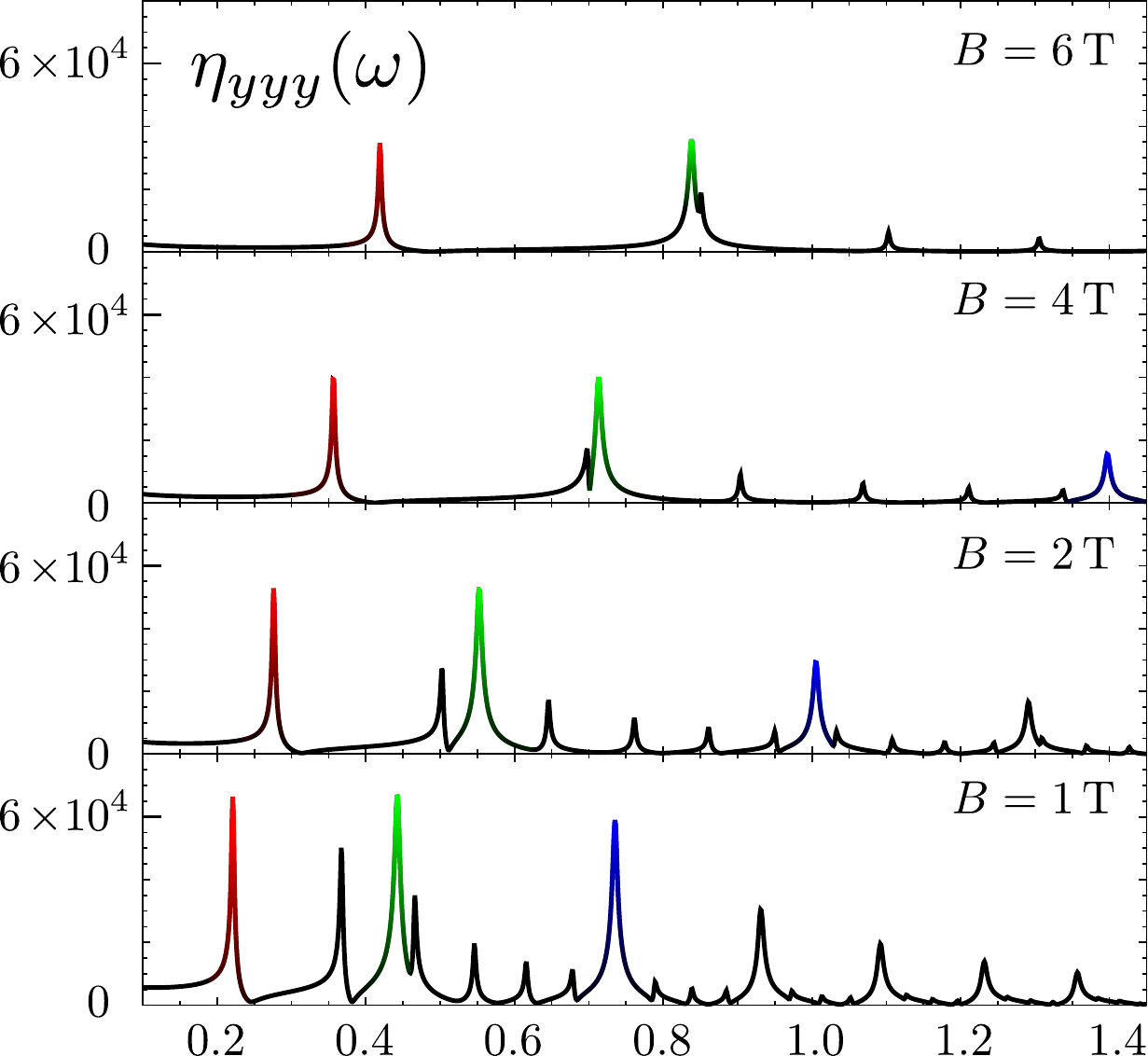}~
\caption{The absolute value of $\eta_{yyy}(\omega)$ (in a.u.) for the case of Gr/SiC under the effect of the magnetic field $B=1\,\text{T}, 2\,\text{T}, 4\,\text{T},~\text{and}~6\,\text{T}$. Colors serve as guides to the eye and depict resonances on the same Landau levels at different values of the magnetic field.}
\label{fig:SHGmag}
\end{figure}

In the presence of the magnetic field, the translational symmetry of the initial problem is explicitly broken. However, the symmetry with respect to inversion in $\kv$-space in the Dirac model with the trigonal warping is still preserved. This ensures that the contribution from the triangular diagram $\Pi^{(3)}(\omega)$ to the SHG response remains zero. The explicit calculation of the nonlinear bubble diagram $\Pi^{(2)}(\omega)$ with the above approximations is shown Appendix~\ref{app:magnetic}. The result for the corresponding conversion efficiency for the case of Gr/SiC in the presence of the magnetic field is shown in Fig.~\ref{fig:SHGmag}. Here, we clearly see the multiple-peak structure of the SHG response function due to excitations between Landau levels. A similar picture has been observed experimentally in~\cite{SHGGaAs} for the SHG in GaAs material. It can be shown that a selection rule for the allowed transitions between Landau levels is $\Delta n = \pm 1$. A more precise analysis of Fig.~\ref{fig:SHGmag} allows to distinguish two types of peaks with different intensities that correspond to the contribution of different diagrams ``$c$'' and ``$d$'' shown in Fig.~\ref{eq:diagrams} to the SHG.

As we observe above, the SHG signal from the Gr/hBN in the absence of the magnetic field is much smaller than the signal coming from the clean hBN at experimentally accessible frequencies. An application of the homogeneous magnetic field results in the formation of Landau level, which in the case of small value of the effective mass of the graphene-based heterostructures appear already at energies below the band-gap of the insulating substrate.
The resonance on the Landau levels drastically enhances the SHG signal from the graphene placed on top of the hBN, while the SHG responce from the hBN remains off-resonant at energies below the bang-gap. Therefore, the inclusion of the magnetic field simplifies the detection of the SHG signal from the graphene flake disposed on top of the insulating substrate.

\section{Conclusion}

In this work we present a consistent calculation of the SHG response from the graphene-based heterostructures. In our calculations we first start with the full tight-binding dispersion and obtain the nonlinear optical spectrum for experimentally relevant frequencies of the applied light. We find that the conversion efficiency has three pairs of resonances that correspond to optical excitations between the band-gap, van Hove singularity and band width. We also observe that the off-resonant behavior of the SHG response function linearly depends on the mass term, contrary to what has been reported in previous studies. The problem of distinguishing the small signal from the graphene on top of the insulating substrate is proposed to resolve here by the inclusion of the magnetic field. The presence of the latter in the system leads to a formation of Landau levels in the energy spectrum at energies below the band-gap of the insulating substrate. This allows to obtain resonant SHG excitations for the considered heterostructure keeping the SHG signal from hBN off-resonant. The magnetic field is included in the theory analytically in the framework of the simplified Dirac model with the trigonal warping. The limit of applicability of this approximation is carefully discussed. For instance, we find that the derived approximation is valid at energies below 2\,eV, which is sufficient for a description of the SHG experiment with experimentally relevant frequency of the applied light.

\begin{acknowledgments}
We would like to thank Clement Dutreix, Sergey Semin, Alexey Kimel, and Franca Manghi for fruitful discussions. M.I.K. acknowledges the supported by NWO via Spinoza Prize. The work of E.A.S. was supported by the Russian Science Foundation, Grant 17-72-20041.
\end{acknowledgments}

%\clearpage

%\bibliographystyle{apsrev4-1}
\bibliography{example.bib}

\clearpage
\appendix
\onecolumngrid

\section{Derivation of vertex functions}
\label{app:diagrams}

Explicit relation for the Fourier transform of the nearest-neighbor $t$ and next-nearest-neighbor $t'$ hopping processes in the case of a hexagonal lattice is following
\begin{align}
S_{\bf k} &= ~~t\, \left[\exp(ik_y) + 2 \exp\left(-\frac{ik_y}{2}\right) \cos\left(\frac{\sqrt{3}}{2} k_x\right)\right],\label{Sk}\\
f_{\bf k} &= 2 t'\left[\cos(\sqrt{3} k_x) + 2 \cos\left(\frac{\sqrt{3}}{2} k_x\right)  \cos\left(\frac{3}{2} k_y\right)\right].
\end{align}
The expression for velocities can be derived following the Ref.~\onlinecite{valleypol}
\begin{equation}
{\rm v}^{(1)\,\sigma\sigma'}_{\alpha}({\bf k}) = \frac{1}{e}\left.\frac{\delta H^{\sigma\sigma'}_{\bf k}}{\delta A_\alpha}\right\rvert_{A=0} = \partial_{k_\alpha}H^{\sigma\sigma'}_{\bf k} - i(r_\alpha^{\sigma'}-r_\alpha^{\sigma})H^{\sigma\sigma'}_{\bf k},
\end{equation}
where ${\bf A}$ is the vector potential of the applied light introduced by the Peierls substitution, and $\sigma, \sigma'$ indicate pseudospin degrees of freedom related to a sublattice space. $r$ indicates the atomic position within the unit cell.
%Specializing to the case we are interested in, we can calculate the expression for Hamiltonian \eqref{hexham} with dispersion ????.
%\eqref{hexdisp}.
The resulting expression for velocities becomes
\begin{align}
{\rm v}^{(1)}_x(\textbf{k}) &=  \left(\begin{matrix}
{\rm v}_x^{(1)\,AA}(\textbf{k}) & -\sqrt{3}ta\, e^{\frac{-i k_y a}{2}} \sin\left(\frac{\sqrt{3} k_xa}{2}\right)\\
-\sqrt{3}ta\, e^{\frac{+i k_y a}{2}} \sin\left(\frac{\sqrt{3} k_xa}{2}\right) & {\rm v}_x^{(1)\,BB}(\textbf{k}) \\
\end{matrix}\right), \\
{\rm v}^{(1)}_y(\textbf{k}) &= \left(\begin{matrix}
-6t'a\,\cos(\frac{\sqrt{3}}{2}k_xa) \sin(\frac{3}{2}k_ya) & -3ita\, e^{ -\frac{ik_ya}{2}} \cos\left(\frac{\sqrt{3}}{2} k_xa\right) \\
3ita\, e^{ \frac{ik_ya}{2}} \cos\left(\frac{\sqrt{3}}{2} k_xa\right) & -6t'a\,\cos(\frac{\sqrt{3}}{2}k_xa) \sin(\frac{3}{2}k_ya) \\
\end{matrix}\right),
\end{align}
where
${\rm v}_x^{(1)\,AA}(\kv)={\rm v}_x^{(1)\,BB}(\kv)= -2\sqrt{3}t'a\left(\sin(\sqrt{3}k_xa) + \sin(\frac{\sqrt{3}}{2}k_xa) \cos(\frac{3}{2}k_ya) \right)$.
In the same way, recalling the equation for the two-photon velocity
\begin{align}
&{\rm v}^{(2)\,\sigma\sigma'}_{\alpha\beta} ({\bf k})= \frac{1}{e^2}\left.\frac{\delta H^{\sigma\sigma'}_{\bf k}}{\delta A_\alpha \delta A_\beta}\right\rvert_{A=0} = \left[\partial_{k_\alpha}\partial_{k_\beta}+i(\rho_\alpha^{\sigma'}-\rho_\alpha^{\sigma})\partial_{k_\beta}+i(\rho_\beta^{\sigma'}-\rho_\beta^{\sigma})\partial_{k_\alpha}-(\rho_\beta^{\sigma'}-\rho_\beta^{\sigma})(\rho_\alpha^{\sigma'}-\rho_\alpha^{\sigma})\right]H^{\sigma\sigma'}_{\bf k}. \notag
\end{align}
This leads to the following expressions for the components of the second-order velocities
\begin{align}
{\rm v}^{(2)}_{xx}({\bf k})&=\left(
\begin{array}{cc}
{{\rm v}_{xx}^{(2)\,AA}}(\kv) & -\frac{3ta^2}{2}  e^{-\frac{i k_ya}{2}}  \cos \left(\frac{\sqrt{3} k_xa}{2}\right) \\
-\frac{3ta^2}{2}  e^{\frac{i k_ya}{2}}  \cos \left(\frac{\sqrt{3} k_xa}{2}\right) & {{\rm v}_{xx}^{(2)\,BB}}(\kv) \\
\end{array}
\right),\\
{\rm v}^{(2)}_{yy}({\bf k})&=\left(
\begin{array}{cc}
-9 t'a^2 \cos \left(\frac{\sqrt{3} k_x}{2}\right) \cos \left(\frac{3 k_y}{2}\right) & -\frac{1}{2} e^{-\frac{1}{2} \left(i k_y\right)} t \left(\cos \left(\frac{\sqrt{3} k_x}{2}\right)+8 e^{\frac{3 i k_y}{2}}\right) \\
 -\frac{1}{2} e^{-i k_y} t \left(e^{\frac{3 i k_y}{2}} \cos \left(\frac{\sqrt{3} k_x}{2}\right)+8\right) & -9 t'a^2 \cos \left(\frac{\sqrt{3} k_x}{2}\right) \cos \left(\frac{3 k_y}{2}\right) \\
\end{array}
\right), \notag
\end{align}
where
${{\rm v}_{xx}^{(2)\,AA}}(\kv) = {{\rm v}_{xx}^{(2)\,BB}}(\kv) = -3 t'a^2 \left(\cos \left(\frac{\sqrt{3} k_xa}{2}\right) \cos \left(\frac{3 k_ya}{2}\right)+2 \cos \left(\sqrt{3} k_xa\right)\right)$, which is the expression for the two-electrons-two-photons vertex.

%In our case we are interested in diagonal elements of the three-momenta matrix element $\Omega_{\alpha\beta\gamma}$ for symmetry reasons (see \cite{marg}).

\section{Evaluation of diagrams for the SHG}
\label{app:tensor}

An important step in our derivation is to take explicitly the summation over internal frequencies in the loop of equation \ref{diag}. This simplifies considerably the expressions to be computed and gives physical insight into the problem, allowing to show explicitly the connection between formulas and physical transitions.
The Green's function $G({\bf k}, \nu)$ of the initial problem~\eqref{hexham} is
\begin{align}
%& G(\textbf{k},\nu) = \frac{1}{\mathbb{1}(i\nu + \mu - f_{\bf k})- \hat{H}_{\bf k}} = \frac{\mathbb{1}(i\nu + \mu-f_{\bf k}) + \boldsymbol{\sigma} \cdot \boldsymbol{\xi}_{\bf k}}{(i \nu + \mu - f^{\phantom{1}}_{\bf k})^2 - \xi_{\bf k}^{2}}.\\
\hat{G}(\textbf{k},\nu) = \frac{1}{\mathbb{1}(i\nu + \mu)- \hat{H}_{\bf k}} = \frac{\mathbb{1}(i\nu + \mu-f_{\bf k}) + \boldsymbol{\hat\sigma} \cdot \boldsymbol{\xi}_{\bf k}}{(i \nu + \mu - f^{\phantom{1}}_{\bf k})^2 - \xi_{\bf k}^{2}}, \label{eq:GF}
\end{align}
where $\mathbb{1}$ is the identity matrix and  $\boldsymbol{\hat\sigma}=(\hat\sigma_x,\hat\sigma_y,\hat\sigma_z)$ is the vector of Pauli Matrices in the sublattice space. We also define $\boldsymbol{\xi}_{\bf k} = (\mathrm{Re}\, S_\kv,\mathrm{Im}\, S_\kv,m)$ and $\xi_{\rm k} = \sqrt{|S^{\phantom{|}}_{\bf k}|^2+m^2}$.
%with $\textbf{S}_k$, $\xi_k$ and $m$ defined above and $\mu$ the chemical potential.

This representation of the Green's function is not convenient to look for a compact expression for the diagram~\eqref{diag}, so we define the so-called \emph{spectral representation} of the Green's function (see e.g.\cite{le2000thermal}). It's an expansion of the denominator in the above expression in simple fractions
\begin{align}
\frac{\mathbb{1}(i\nu + \mu -f_{\bf k}) + \boldsymbol{\hat\sigma} \cdot \boldsymbol{\xi}_{\bf k}}{(i \nu + \mu - f^{\phantom{1}}_{\bf k})^2 - \xi_{\bf k}^{2}} = \frac{\hat{A}}{i \nu + \mu - f_{\bf k} - \xi_{\bf k}} + \frac{\hat{B}}{i \nu + \mu - f_{\bf k} + \xi_{\bf k}}.
\end{align}
%The system obtained equating coefficients of the same powers of the frequency reads:
%\begin{align}
%\left\lbrace
%\begin{matrix}
%\hat{A} + \hat{B} = \mathbb{1} \\
%\hat{A} \xi_\textbf{k} - \hat{B} \xi_\textbf{k} = \boldsymbol{\sigma} \cdot \boldsymbol{\xi}_{\bf k}
%\end{matrix}
%\right.
%\end{align}
Solving for the matrix coefficients $\hat{A}$ and $\hat{B}$, we find the following expression for the propagator
\begin{equation}
\hat{G}({\bf k},\nu) = \sum\limits_{s=\pm 1}  \frac{\hat{\Lambda}^{s}(\textbf{k})}{i\nu + \mu - f_{\bf k} - s\xi_{\bf k}},
\end{equation}
where we have introduced projectors over positive and negative energy states
\begin{align}
\hat{\Lambda}^{s}(\textbf{k}) =  \frac{1}{2}\left(\mathbb{1} + \frac{s}{\xi_{\bf k}}\,\boldsymbol{\hat\sigma} \cdot \boldsymbol{\xi}_{\bf k}\right).
\end{align}
%The generalization of the method to systems without particle-hole symmetry and to the multiband case is straightforward at this level, and it amounts to consider the replacement $s\,\xi^{\phantom{s}}_{\bf k} \rightarrow \xi_{\bf k}^s$, where now ``$s$'' plays the role of a spin-band index.
We now define the third-order factor for the diagram as
\begin{equation}
\Omega^{ss's''}_{\alpha \beta \gamma}(\textbf{k}) = \Tr\left[ \hat{\rm v}^{(1)}_{\alpha} \hat{\Lambda}^{s}(\textbf{k}) \hat{\rm v}^{(1)}_{\beta} \hat{\Lambda}^{s'}(\textbf{k}) \hat{\rm v}^{(1)}_{\gamma} \hat{\Lambda}^{s''}(\textbf{k})\right],
\end{equation}
Most importantly, this object does not depend on the frequency $\nu$.
Omitting the integration over momenta in~\eqref{diag}, one gets
\begin{equation}
\Pi^{(3)}_{\alpha\beta\gamma}(\omega,\kv) = \sum\limits_{\nu,s,s',s''} \frac{\Omega^{ss's''}_{\alpha \beta \gamma}(\textbf{k})}{[i\nu +  \mu  - f^{\phantom{s}}_{{\bf k}}-\xi^{s}_{{\bf k}}] [i\nu + i\omega + \mu - f^{\phantom{s}}_{{\bf k}}-\xi^{s'}_{{\bf k}}] [i\nu - i\omega+ \mu - f^{\phantom{s}}_{{\bf k}}-\xi^{s''}_{{\bf k}}]}.
\label{beforesum}
\end{equation}
%with $s,s^{\prime},s^{\prime\prime}=\pm 1$.
%Repeating what is done in Le Bellac for scalar fields, we can manipulate this expression to obtain something a lot easier.
Now we concentrate on the following part of the denominator and we notice that it can be conveniently manipulated exploiting partial fractions
\begin{align}
&\frac{1}{\left(i\nu + i\omega +\mu - f^{\phantom{s}}_{{\bf k}}-\xi^{s'}_{{\bf k}}\right) \left(i\nu - i\omega + \mu - f^{\phantom{s}}_{{\bf k}}-\xi^{s''}_{{\bf k}}\right)} =
\frac{A}{i\nu + i\omega +\mu - f^{\phantom{s}}_{{\bf k}}-\xi^{s'}_{{\bf k}}} + \frac{B}{   i\nu - i\omega + \mu - f^{\phantom{s}}_{{\bf k}}-\xi^{s''}_{{\bf k}}}, \notag
\end{align}
where
\begin{align}
A = -B = -\frac{1}{   2i\omega - \left(\xi^{s'}_{{\bf k}}-\xi^{s''}_{{\bf k}}\right)}.\notag
\end{align}
The useful fact is that the previously complicated evaluation is reduced to the evaluation of two polarization bubbles. The expression for this diagram at finite chemical potential can be found, for example, in ~\cite{le2000thermal} (p. 157).
In our case one gets
\begin{align}
\label{Pi3}
\Pi^{(3)}_{\alpha\beta\gamma}(\kv,\omega) =  \sum\limits_{s,s',s''}\frac{-\Omega^{ss's''}_{\alpha \beta \gamma}(\textbf{k})}{   2i\omega - \left(\xi^{s'}_{{\bf k}}-\xi^{s''}_{{\bf k}}\right)} \sum\limits_{\nu}  \frac{1}{[i\nu  + \mu - f^{\phantom{s}}_{{\bf k}}-\xi^{s}_{{\bf k}}]}\left[ \frac{1}{i\nu + i\omega +\mu - f^{\phantom{s}}_{{\bf k}}-\xi^{s'}_{{\bf k}}} - \frac{1}{ i\nu - i\omega + \mu - f^{\phantom{s}}_{{\bf k}}-\xi^{s''}_{{\bf k}}} \right].
\end{align}
The first term in the second line of Eq.~\ref{Pi3} reads
\begin{align}
\label{bubble1}
&\sum\limits_{\nu}\frac{1}{ [i\nu  + \mu - f^{\phantom{s}}_{{\bf k}}-\xi^{s}_{{\bf k}}] [i\nu + i\omega + \mu - f^{\phantom{s}}_{{\bf k}} - \xi^{s'}_{{\bf k}}]} =
\frac{1}{i\omega + \left( \xi^{s}_{{\bf k}} - \xi^{s'}_{{\bf k}}\right)} \sum\limits_{\nu}   \left[ \frac{1}{ i\nu + \mu - f^{\phantom{s}}_{{\bf k}} - \xi^{s}_{{\bf k}}} - \frac{1}{ i\nu+i\omega+\mu-f^{\phantom{s'}}_{{\bf k}} - \xi^{s'}_{{\bf k}}}\right].
\end{align}
Now we see that this expression can be summed introducing a convergence factor  $e^{i\nu_n\eta}$ with $\eta \rightarrow 0$ in every term (see e.g.~\cite{fetter}, p. 272). Then, the evaluation of the Matsubara sum becomes simply the evaluation of the function $\frac{1}{2} -  n_F(\xi)$ in correspondence of the poles of the function involved, where $n_{\rm F}(\xi-\mu) = \left(e^{\beta(\xi-\mu)}+1\right)^{-1}$ is the Fermi distribution function. In this case there are two poles: $i \nu = -\mu + f^{\phantom{s}}_{\bf k}+\xi^{s}_{\bf k}$ for the first fraction in the square brackets and $i \nu+\omega = -\mu + f^{\phantom{s}}_{{\bf k}} + \xi^{s'}_{{\bf k}}$ for the second one, keeping in mind that $i \omega$ is a bosonic Matsubara's frequency and the exponential of bosonic frequencies gives just a factor 1.
We can then derive that the sum over frequencies as
\begin{align}
\label{part}
&\sum\limits_{\nu}\frac{1}{ [i\nu  + \mu - f^{\phantom{s}}_{{\bf k}}-\xi^{s}_{{\bf k}}] [i\nu + i\omega + \mu - f^{\phantom{s}}_{{\bf k}} - \xi^{s'}_{{\bf k}}]} =
\frac{1}{i\omega + \left( \xi^{s}_{{\bf k}} - \xi^{s'}_{{\bf k}}\right)} \left[n_{\rm F}\left(\xi^{s}_{{\bf k}} + f^{\phantom{s}}_{{\bf k}}-\mu\right) - n_{\rm F}\left(\xi^{s'}_{{\bf k}} + f^{\phantom{s'}}_{{\bf k}}-\mu\right) \right].
\end{align}
The second term in the second line of Eq.~\ref{Pi3} is given by a similar expression with the replacement $s' \to s''$.
%We have of course used a mathematically not well defined way to take the summation over Matsubara's.
The full expression for the diagram can be obtained putting these two terms together and reads

\begin{align}
\label{eq:PiSHGapp}
&\Pi^{(3)}_{\alpha\beta\gamma}({\bf k},\omega) = \sum\limits_{s,s',s''}\frac{-\Omega^{ss's''}_{\alpha \beta \gamma}(\textbf{k})}{   2i\omega -\left(\xi^{s'}_{{\bf k}}-\xi^{s''}_{{\bf k}}\right)}
\left[ \frac{n_{\rm F}\left(\xi^{s'}_{{\bf k}} + f^{\phantom{s}}_{{\bf k}}-\mu\right) - n_{\rm F}\left(\xi^{s}_{{\bf k}} + f^{\phantom{s'}}_{{\bf k}}-\mu\right)}{i\omega - \left( \xi^{s'}_{{\bf k}} - \xi^{s}_{{\bf k}}\right)} -\frac{n_{\rm F}\left(\xi^{s}_{{\bf k}} + f^{\phantom{s}}_{{\bf k}}-\mu\right) - n_{\rm F}\left(\xi^{s''}_{{\bf k}} + f^{\phantom{s''}}_{{\bf k}}-\mu\right)}{i\omega  - \left( \xi^{s}_{{\bf k}} - \xi^{s''}_{{\bf k}}\right)} \right].
\end{align}
%where $n_{\rm F}(x) = \left(e^{\,\beta x}+1\right)^{-1}$ is the Fermi distribution function and
%\begin{align}
%\Omega^{ss's''}_{\alpha \beta \gamma}(\textbf{k}) &= \Tr\left[ \text{v}^{(1)}_{\alpha} \Lambda^{s}(\textbf{k}) \text{v}^{(1)}_{\beta} \Lambda^{s'}(\textbf{k}) \text{v}^{(1)}_{\gamma} \Lambda^{s''}(\textbf{k})\right],
%\end{align}

Now we take into account that in our case the $s$ indexes can assume only values $\pm 1$.
The result becomes different from 0 only if $s \neq s'$ or/and $s \neq s''$, otherwise one of the numerators in brackets in Eq.~\ref{eq:PiSHG} vanishes.
Therefore, one gets three contributions
\begin{align}
&1) \,\,  \sum_{s=\pm1} \frac{\Omega^{s,s,-s}_{\alpha \beta \gamma}(\textbf{k})}{2(i\omega - s \xi_{\bf k})} \left[\frac{n_{\rm F}\left(s\xi_{\bf k} + f_{\bf k}-\mu\right) - n_{\rm F}\left(-s\xi_{\bf k} + f_{\bf k}-\mu\right)}{i\omega - 2s\xi_{\bf k}} \right], \label{first}\\
&2) \,\,  \sum_{s=\pm1} \frac{\Omega^{s,-s,s}_{\alpha \beta \gamma}(\textbf{k})}{2(i\omega + s \xi_{\bf k})} \left[\frac{n_{\rm F}\left(s\xi_{\bf k} + f_{\bf k}-\mu\right) - n_{\rm F}\left(-s\xi_{\bf k} + f_{\bf k}-\mu\right)}{i\omega + 2s\xi_{\bf k}} \right], \label{second}\\
&3) \,\,  \sum_{s=\pm1} \frac{-\Omega^{s,-s,-s}_{\alpha \beta \gamma}(\textbf{k})}{\omega^2 + 4 \xi^2_{\bf k}}\left[n_{\rm F}\left(s\xi_{\bf k} + f_{\bf k}-\mu\right) - n_{\rm F}\left(-s\xi_{\bf k} + f_{\bf k}-\mu\right)\right]. \label{third}
\end{align}
Rearranging terms in \eqref{first} and \eqref{second}, and changing $s$ to $-s$ in the second one, we get the expression
\begin{align}
\Pi^{(3)}_{\alpha\beta\gamma}({\bf k},\omega) &= \sum_{s=\pm1} \frac{\Omega^{s,s,-s}_{\alpha \beta \gamma}(\textbf{k})-\Omega^{-s,s,-s}_{\alpha \beta \gamma}(\textbf{k})}{2(i\omega - s \xi_{\bf k})} \left[\frac{n_{\rm F}\left(s\xi_{\bf k} + f_{\bf k}-\mu\right) - n_{\rm F}\left(-s\xi_{\bf k} + f_{\bf k}-\mu\right)}{i\omega - 2s\xi_{\bf k}} \right] \label{eq:PiSHG}\\
&-\sum_{s=\pm1} \frac{\Omega^{s,-s,-s}_{\alpha \beta \gamma}(\textbf{k})}{\omega^2 + 4 \xi^2_{\bf k}}\left[n_{\rm F}\left(s\xi_{\bf k} + f_{\bf k}-\mu\right) - n_{\rm F}\left(-s\xi_{\bf k} + f_{\bf k}-\mu\right)\right].\notag
\end{align}

\begin{figure}[t!]
\includegraphics[width=0.47\linewidth]{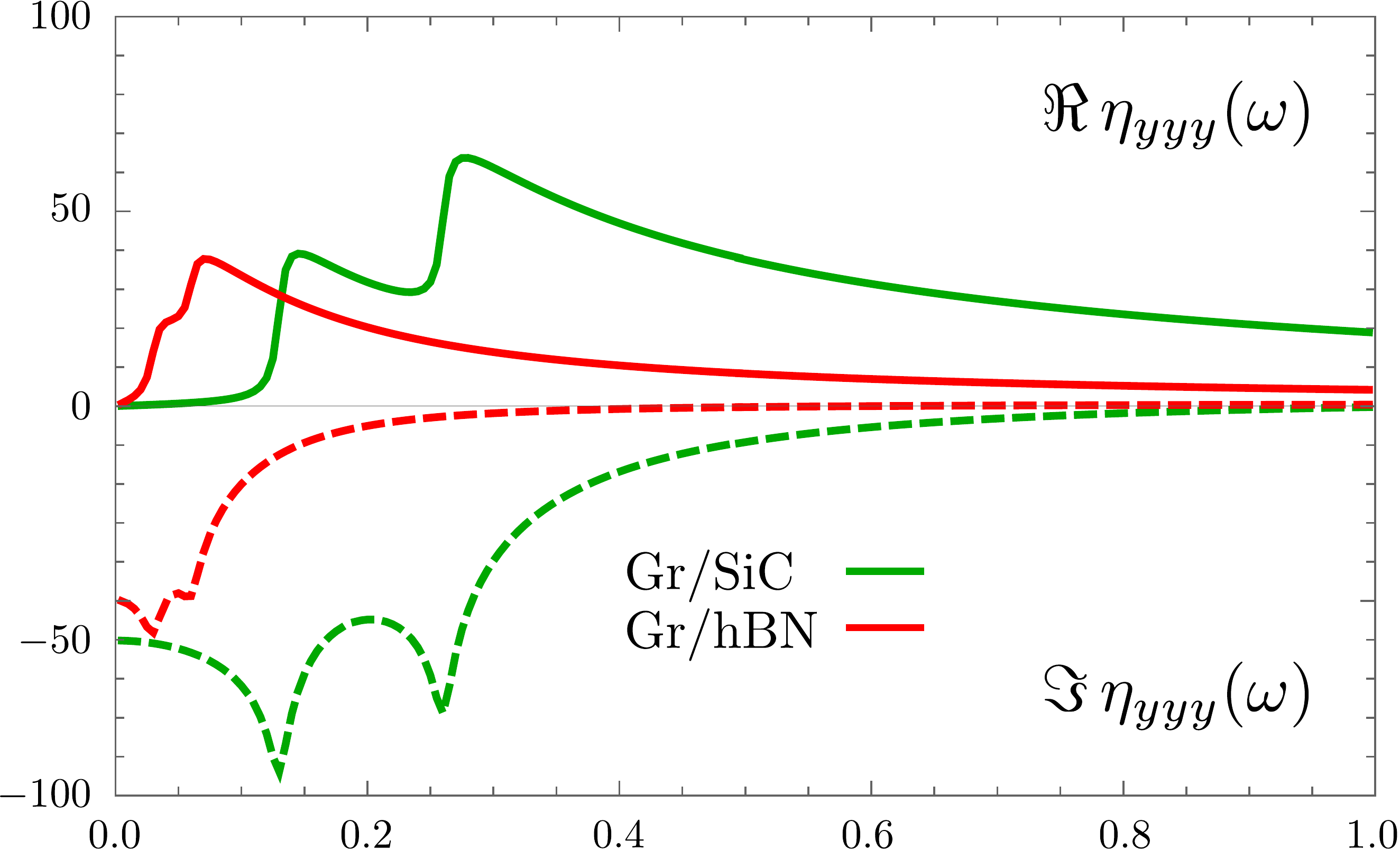} ~~~
\caption{\label{fig:PiX} Real (solid line) and Imaginary (dashed line) part of the conversion efficiency $\eta_{yyy}(\omega)$ for Gr/SiC (green color) and Gr/hBN (red color).}
\end{figure}

The next step is to calculate the contribution coming from the non-linear bubble.
The evaluation of the Matsubara summation can be done in the same way as we did for the triangular diagram.
The integrand of the non-linear bubble $\Pi^{(2)}_{\alpha\beta\gamma}(\omega)$ becomes:
\begin{align}
&\Pi^{(2)}_{\alpha\beta\gamma}({\bf k}, \omega) = \sum_{s,s'}\Theta^{ss'}_{\alpha\beta\gamma}({\bf k})\sum\limits_{\nu}\left( \frac{1}{i\nu-i\omega+\mu-f_{\bf k}-\xi^{s}_{\bf k}} \cdot\frac{1}{i\nu+i\omega+\mu-f_{\bf k}-\xi^{s'}_{\bf k}}+ \frac{2}{i\nu+\mu-f_{\bf k}-\xi^{s}_{\bf k}} \cdot\frac{1}{i\nu+i\omega+\mu-f_{\bf k}-\xi^{s'}_{\bf k}}\right),
\end{align}
where
\begin{align}
\Theta^{ss'}_{\alpha\beta\gamma}(\textbf{k} ) = \Tr\left[\hat{\rm v}^{(2)}_{\alpha\beta} \hat{\Lambda}^s(\textbf{k}) \hat{\rm v}^{(1)}_\delta \hat{\Lambda}^{s'}(\textbf{k})\right].
\end{align}
We notice that this expression is identical to Eq.~\ref{part} except for $-i\omega$ in the first factor. Adapting Eq.~\ref{part}, we can then easily derive the expression for the bubble to be
\begin{equation}
\Pi^{(2)}_{\alpha\beta\gamma}({\bf k},\omega) = \sum_{s,s'}\Theta^{ss'}_{\alpha\beta\gamma}({\bf k}) \left[\frac{n_{\rm F}(\xi^{s'}_{\bf k}+f_{\bf k}-\mu)-n_{\rm F}(\xi^{s}_{\bf k}+f_{\bf k}-\mu)}{2i\omega-\left(\xi^{s'}_{{\bf k}}-\xi^{s}_{{\bf k}}\right)}+2 \, \frac{n_{\rm F}(\xi^{s'}_{\bf k}+f_{\bf k}-\mu)-n_{\rm F}(\xi^{s}_{\bf k}+f_{\bf k}-\mu)}{i\omega-\left(\xi^{s'}_{{\bf k}}-\xi^{s}_{{\bf k}}\right)}\right].
\label{nonlinbubbleSHGapp}
\end{equation}
As in the previous case, the only possible contribution is given by $s \neq s'$, so we can write
\begin{equation}
\Pi^{(2)}_{\alpha\beta\gamma}({\bf k},\omega) = \sum_{s}\Theta^{-s,s}_{\alpha\beta\gamma}({\bf k}) \left[\frac{n_{\rm F}(s\xi_{\bf k}+f_{\bf k}-\mu)-n_{\rm F}(-s\xi_{\bf k}+f_{\bf k}-\mu)}{2i\omega-2s\xi_{{\bf k}}}+2 \, \frac{n_{\rm F}(s\xi_{\bf k}+f_{\bf k}-\mu)-n_{\rm F}(-s\xi_{\bf k}+f_{\bf k}-\mu)}{i\omega-2s\xi_{{\bf k}}}\right].
\label{nonlinbubbleSHGappspec}
\end{equation}
Results for the Real and Imaginary parts of the conversion efficiency $\eta_{yyy}(\omega)\sim\Pi^{(2)}_{yyy}(\omega)/\omega$ for real frequencies is shown in Fig.~\ref{fig:PiX}.
%In this study we did not explicitly consider the case where inversion symmetry in the Brillouin zone is broken. However this can be directly included by just introducing a momentum dependent chemical potential $\mu = \mu(\kv)$. In that case we expect the triangular diagram to be non-zero.

\section{SHG in Dirac model with trigonal warping}
\label{app:Dirac}

At low energies and the initial Hamiltonian matrix $\hat{H}_{\kv}$ can be expanded around $K$ and $K'$ points with respect to momentum small momentum $\kv$. Then, we get a Dirac approximation with the trigonal warping for the initial problem
\begin{align}
\hat{H} = v\left[\tau k_x \hat{\sigma}_x + k_y\hat{\sigma}_y\right] + m\hat{\sigma}_{z} + \lambda [2\tau \hat{\sigma}_y k_x k_y - \hat{\sigma}_x  (k_x^2 -k_y^2)],
\end{align}
where we have introduced the electron speed at the conical points $v = \frac{3at}{2}$, valley index $\tau = \pm1$, and the trigonal warping parameter $\lambda=3a^2t/8$. Velocities in this case are defined in the low energy limit (continuous limit) as
\begin{equation}
\hat{\rm v}^{(1)}_{\alpha} = \frac{1}{e}\left.\frac{\delta \hat{H}}{\delta A_\alpha }\right\rvert_{A=0} = \left.\partial_{k_\alpha} \hat{H}^{\phantom{s}}_{\phantom{s}} \right\rvert_{A=0}, ~~~
\hat{\rm v}^{(2)}_{\alpha\beta} = \frac{1}{e^2}\left.\frac{\delta^2 \hat{H}}{\delta A_\alpha \delta A_\beta }\right\rvert_{A=0}=\left.\partial_{k_\alpha}  \partial_{k_\beta}  \hat{H}^{\phantom{s}}_{\phantom{s}}\right\rvert_{A=0}.
\end{equation}
Here, $\hat{\rm v}^{(1)}$, $\hat{\rm v}^{(2)}$ and $\hat{H}$ are $2 \times 2$  matrices in the sublattice space.

We use the simplest approximation, that amounts to include the contribution of the trigonal warping only in the non-linear velocity ${\rm v}^{(2)}_{\alpha\beta}$. This is justified by the fact that in the dispersion relation the trigonal warping parameter, which is already smaller than the electronic speed, appears multiplied by the squared momentum. So, the corresponding correction to the position of the poles of the Green's function is assumed to be small. On the other hand, it is the first non-zero contribution to the non-linear velocity, so it cannot be excluded from the consideration. This amounts to have a bubble diagram with a usual velocity ${\rm v}^{(1)}$ that does not depend on $\lambda$, and a non-linear vertex that is linear in $\lambda$. This is the lowest order in $\lambda$ that describes SHG. In this case the velocities can be obtained differentiating the Hamiltonian with respect to momentum as discussed above. So, they are simply proportional to Pauli Matrices as follows.
\begin{align}
\hat{{\rm v}}^{(1)}_x = v \tau \hat{\sigma}_x; ~~~
\hat{{\rm v}}^{(1)}_{y} = v \hat{\sigma}_y; ~~~
\hat{{\rm v}}^{(2)}_{xx}=\hat{{\rm v}}^{(2)}_{yy} = - 2\lambda \hat{\sigma}_x; ~~~
\hat{{\rm v}}^{(2)}_{xy} = \hat{{\rm v}}^{(2)}_{yx}= 2\tau\lambda  \hat{\sigma}_y.
\label{magveloc}
\end{align}
We take a further approximation and consider only first order terms in trigonal warping in the diagram $\Pi^{(2)}(\omega)$. Since the $\hat{{\rm v}}^{(2)}_{yy}(\textbf{k})$ is already proportional to the trigonal warping parameter, we neglect the contribution of the trigonal warping in the Green's function. As we discuss in the main text (see Fig.~\ref{fig:comparisonDir}), the approximated result quantitatively agrees with the one of the tight-binding model.
The bubble diagram $\Pi^{(2)}$ in this case can be expressed in particularly simple way. Starting from Eq. \eqref{nonlinbubbleSHGappspec}, we can considerably simplify this expression in our Dirac approximation.
Starting from
\begin{align}
\Pi^{(2)}_{\alpha\beta\gamma}(\omega) = \sum\limits_{|\kv|< k_c}\sum\limits_{\text{cones}}\Pi^{(2)}_{\alpha\beta\gamma}({\bf k},\omega) = 3 \sum\limits_{|\kv|< k_c}\sum\limits_{\tau = \pm 1} \sum_{s}\Theta^{-s,s}_{\alpha\beta\gamma}({\bf k}_\tau) \left[\frac{n_{\rm F}(s\varepsilon_{\kv}-\mu)-n_{\rm F}(-s\varepsilon_{\kv}-\mu)}{2i\omega-2s\varepsilon_{\kv}}+2 \, \frac{n_{\rm F}(s\varepsilon_{\kv}-\mu)-n_{\rm F}(-s\varepsilon_{\kv}-\mu)}{i\omega-2s\varepsilon_{\kv}}\right],
\end{align}
where we consider that there are three couples of $K$ and $K'$ points in the Brillouin zone, $\tau$ is the valley index and $k_c$ is the maximum value of the momentum for which the Dirac approximation is valid.
Now we notice that every component where $\tau$ enters only once, is zero when summed over valley index.
The only non-zero contributions are therefore those involving ${\rm v}^{(1)}_x$ and ${\rm v}^{(2)}_{xy/yx}$ or
${\rm v}^{(1)}_y$ and ${\rm v}^{(2)}_{xx/yy}$, as discussed in the main text with symmetry considerations.
It is straightforward to show that this considerations are all equal to each other up to minus sign.
We can then evaluate the following quantity
\begin{align}
\Theta^{-s,s}({\bf k})=\sum\limits_{\tau=\pm 1}\Theta^{-s,s}_{yyy}({\bf k}_\tau)=2 \Tr\left[\hat{\rm v}^{(2)}_{yy} \hat{\Lambda}^s(\textbf{k}) \hat{\rm v}^{(1)}_y \hat{\Lambda}^{s'}(\textbf{k})\right]= \frac{4 i \lambda \, v \, m \, s}{\varepsilon_{\kv}} \end{align}
Let us also recall that
$ \,\,n_{\rm F}(s\varepsilon_{\kv}-\mu)-n_{\rm F}(-s\varepsilon_{\kv}-\mu) = s \tanh\left(\frac{\beta \varepsilon_{\kv}}{2}\right)$.
Now we can plug these expressions back to the general expression for $\Pi^{(2)}(\omega)$ and we get
\begin{align}
&\Pi^{(2)}_{\alpha\beta\gamma}(\omega) =  12 i \lambda \, v \, m \sum\limits_{|\kv|< k_c} \sum_{s} \frac{1}{\varepsilon_{\kv}} \tanh\left(\frac{\beta \varepsilon_{\kv}}{2}\right)\left[\frac{1}{2i\omega-2s\varepsilon_{\kv}}+\, \frac{2}{i\omega-2s\varepsilon_{\kv}}\right] =\\& = 12 i \lambda \, v \, m \sum\limits_{|\kv|< k_c} \frac{\tanh\left(\frac{\beta \varepsilon_{\kv}}{2}\right)}{\varepsilon_{\kv}} \left[\frac{1}{2i\omega-2\varepsilon_{\kv}}+\frac{1}{2i\omega+2\varepsilon_{\kv}}+\frac{2}{i\omega-2\varepsilon_{\kv}}+\frac{2}{i\omega+2\varepsilon_{\kv}}\right]=\notag\\&=12 i \lambda \, v \, m \; i\omega \sum\limits_{k< k_c} \frac{\tanh\left(\frac{\beta \varepsilon_{\kv}}{2}\right)}{\varepsilon_{\kv}  }\left[\frac{1}{(i\omega)^2-\varepsilon_{\kv}^2}+\frac{1}{(i\omega/2)^2-\varepsilon_{\kv}^2}\right],
\end{align}
where we integrated over the angular coordinate since there is no angular dependence and thus the last integral is taken over the modulus $k$ of the momentum only.

\section{SHG in the presence of the magnetic field}
\label{app:magnetic}

In order to consider the effect of the magnetic field, we add the a vector potential describing the incident light using the Peierls substitution as
\begin{align}
\hat{H}_\tau= v\hat{\boldsymbol{\sigma}}_\tau \cdot \left(\hat{{\bf p}}+e{\bf A}_B +e{\bf A}^{\text{rad}} \right)  + m\hat{\sigma}_z
\label{}
\end{align}
where ${\bf A}_B=\frac{B}{2}(-x,y,0)$ is the vector potential describing the external constant magnetic field and ${\bf A}^{\text{rad}}$ describes the radiation field of the incident light.
We started the result given in Ref. \cite{Mir} for the Green's function of a Dirac particle in a magnetic field.
In that paper it was demonstrated that the Green's function of the problem can be written as
\begin{align}
    G({\bf r},{\bf r}',\omega) = \texttt{exp}\left\lbrace -i \frac{\Phi({\bf r},{\bf r}')}{\Phi_0}\right\rbrace \tilde{G}({\bf r}-{\bf r}',\omega)
\end{align}
where $\Phi({\bf r},{\bf r}') = \int\limits_{\bf r}^{\bf r'} {\bf A}_{B}({\bf z})  \cdot d{\bf z}$, $ \Phi_0$ is the magnetic flux quantum and $\tilde{G}$ is only a function of the difference between $\bf r$ and $\bf r'$.
If we consider the bubble diagram between two points in real space
$\Pi^{(2)}({\bf r},{\bf r}',\omega)  \sim G({\bf r},{\bf r}',\omega) G({\bf r'},{\bf r},-\omega)$, we see that the two phases acquired along the paths are equal but opposite in sign, therefore they cancel each other.
\begin{align}
 \Pi^{(2)}({\bf r},{\bf r}',\omega)  = \sum_\nu \Tr\left[ \hat{{\rm v}}^{(2)}_{yy}\tilde{G}({\bf r}-{\bf r}',\nu+\omega) \hat{{\rm v}}^{(1)}_{\gamma}\tilde{G}({\bf r'}-{\bf r},\nu-\omega) \right] + 2\sum_\nu \Tr\left[ \hat{{\rm v}}^{(2)}_{yy}\tilde{G}({\bf r}-{\bf r}',\nu+\omega) \hat{{\rm v}}^{(1)}_{\gamma}\tilde{G}({\bf r'}-{\bf r},\nu) \right] =  \Pi^{(2)}({\bf r}-{\bf r}',\omega)
\end{align}
where the velocities coincide with those calculated in the Dirac case without magnetic field. This means that the resulting bubble depends just on the translationally invariant part of the Green’s function. This allows us to define a momentum dependent Green's function $G({\bf k},\omega)=\sum\limits_{\bf r} \tilde{G}({\bf r},\omega) e^{i\kv \cdot {\bf r}} $ and to calculate the response using equation \eqref{eq:diagrams}.
The expression for $G({\bf k},\omega)$ was derived in Ref. \cite{Mir} and was found to be
\begin{align}
\hat{G}(\kv, \omega) = -\pi \sum_{n=0}^{+\infty} \frac{\hat{D}_{n}(\kv)}{(i\omega+\mu)^2-{\varepsilon_n}^2},
\label{miranskyExp}
\end{align}
where:
\begin{align}
\hat{D}_{n} (\kv)&= -i \exp\left(-\frac{c\kv^2}{|eB|}\right) (-1)^n\left\lbrace(m\hat{\sigma}_z - i \omega \mathbb{1}) \left[[\mathbb{1}-\hat{\sigma}_z]L_n\left(\frac{c\kv^2}{|eB|}\right)-[\mathbb{1}+\hat{\sigma}_z]L_{n-1}\left(\frac{c\kv^2}{|eB|}\right)\right]+4(k_x \hat{\sigma}_x+k_y \hat{\sigma}_y) L_{n-1}^1\left(\frac{c\kv^2}{|eB|}\right)\right\rbrace,
\label{Laguerre}
\end{align}
$L_{n}^{\alpha}(x)$ are generalized Laguerre Polynomials and $\varepsilon_n = \sqrt{m^2+2 \hbar^2 \Omega_c^2 |n|}$ are the discrete Landau levels of the system with cyclotron frequency $\Omega_c$.
The order of magnitude for the cyclotron frequency in Graphene can be estimated replacing the value of the electron velocity $v=\frac{c}{300}$:
\begin{align}
\Omega_c = \sqrt{\frac{2eBv^2}{\hbar c}} \approx 37 \; \; \mathrm{meV} \sqrt{B(\mathrm{Tesla})}.
\label{cyclotron}
\end{align}

We can now see that the denominator has two poles for each frequency. We can rewrite everything in simple fractions in order to obtain a summation over functions with a single pole. To do so we have to identify terms that depend on $\omega$ in the numerator. It is then useful to split the numerator in two parts. Explicitly this reads $ \hat{D}_n = \hat{\Sigma}_{0}(n,\kv) + i\omega\;\hat{\Sigma}_1(n,\kv)  $,
where we have collected the quantities
\begin{align}
\hat{\Sigma}_{0}(n,\kv) &= -i \exp\left(-\frac{c\kv^2}{|eB|}\right) (-1)^n\left[m\hat{\sigma}_z \hat{\Sigma}_{1}(n,\kv)+4(k_x \hat{\sigma}_x+k_y \hat{\sigma}_y) L_{n-1}^1\left(\frac{c\kv^2}{|eB|}\right)\right], \\
\hat{\Sigma}_{1}(n,\kv) &= ~~~i \exp\left(-\frac{c\kv^2}{|eB|}\right) (-1)^n \left[[\mathbb{1}-\hat{\sigma}_z]L_n\left(\frac{c\kv^2}{|eB|}\right)-[\mathbb{1}+\hat{\sigma}_z]L_{n-1}\left(\frac{c\kv^2}{|eB|}\right)\right].
\end{align}
In this case the electron Green's Function becomes
\begin{align}
\hat{G}( \kv,\omega) =\sum_{n=-\infty}^{+\infty}\frac{\hat{\Lambda}_{n}(\kv)}{i\omega-\mathrm{sgn}(n)\varepsilon_n}.
\label{diracmagngreen}
\end{align}

The projectors $\Lambda_{n}$ are obtained solving the equations to reduce Eq. \eqref{miranskyExp} in simple fractions and are
\begin{align}
\hat{\Lambda}_{n}(\kv) = \frac{1}{2}\left(\hat{\Sigma}_1(n,\kv) -i \; \mathrm{sgn}(n) \frac{\hat{\Sigma}_0(n,\kv)}{\varepsilon_n}\right).
\label{magnprojector}
\end{align}

Now only the quantity at the denominator of the Green’s function depends on the Matsubara frequencies and we can repeat exactly the same steps used to derive Eq. as in the case without magnetic field in order to calculate the bubble diagram. In the Dirac approximation, the triangular diagram $\Pi^{(3)} = 0$.
The bubble diagram with magnetic field then reads
\begin{align}
&\Pi_{\alpha\beta\gamma}^{(2)}(\omega) = \sum_{\nu,\kv} \Tr\left[\hat{{\rm v}}^{(2)}_{\alpha\beta} \hat{G}(\nu +\omega \kv) \hat{{\rm v}}^{(1)}_\gamma \hat{G}(\nu-\omega \kv)\right]+2\sum_{\nu,\kv} \Tr\left[\hat{{\rm v}}^{(2)}_{\alpha\beta} \hat{G}(\nu +\omega \kv) \hat{{\rm v}}^{(1)}_\gamma \hat{G}(\nu \kv)\right] \\
&=\sum_{n,n'=-\infty}^{+\infty} \sum_{\nu,\kv}\Theta_{\alpha\beta\gamma}(\kv,n,n')\left[\frac{1 }{\left(i(\nu+\omega)-\mathrm{sgn}(n)\varepsilon_n\right)\left(i(\nu-\omega)-\mathrm{sgn}(n')\varepsilon_{n'}\right)}+\frac{2 }{\left(i(\nu+\omega)-\mathrm{sgn}(n)\varepsilon_n\right)\left(i(\nu-\omega)-\mathrm{sgn}(n')\varepsilon_{n'}\right)}\right] =\notag\\
&=\sum_{\substack{n,n'=-\infty \\ n \neq n', \kv}}^{+\infty} \Theta_{\alpha\beta\gamma}(\kv,n,n') \left[n_{\mathrm{F}}(\varepsilon_n-\mu)-n_{\mathrm{F}}(\varepsilon_{n'}-\mu)\right]\left[\frac{1}{2i\omega-\mathrm{sgn}(n)\varepsilon_n+\mathrm{sgn}(n')\varepsilon_{n'}}+\frac{2}{i\omega-\mathrm{sgn}(n)\varepsilon_n+\mathrm{sgn}(n')\varepsilon_{n'}}\right], \notag
\end{align}
where
\begin{align}
\Theta_{\alpha\beta\gamma}(\kv,n,n')= \Tr\left[\hat{{\rm v}}^{(2)}_{\alpha\beta} \hat{\Lambda}_{n}(\kv) \hat{{\rm v}}^{(1)}_\gamma \hat{\Lambda}_{n'}(\kv)\right],
\end{align}
and we used Eq.~\ref{part} to go from second to third line. The use of this expression reduces the computational effort necessary to compute the polarization that appears in other methods, because the only factor that depends on the momentum is the matrix element. The three-particle correlation function can be written simply as
\begin{align}
&\Pi_{\alpha\beta\gamma}^{(2)}(\omega)=\sum_{\substack{n,n'=-\infty \\ n \neq n'}}^{+\infty} \Theta_{\alpha\beta\gamma}(n,n') \left[n_{\mathrm{F}}(\varepsilon_n-\mu)-n_{\mathrm{F}}(\varepsilon_{n'}-\mu)\right]\left[\frac{1}{2i\omega-\mathrm{sgn}(n)\varepsilon_n+\mathrm{sgn}(n')\varepsilon_{n'}}+\frac{2}{i\omega-\mathrm{sgn}(n)\varepsilon_n+\mathrm{sgn}(n')\varepsilon_{n'}}\right],
\label{SHGmagneticfield0}
\end{align}
where $\Theta_{\alpha\beta\gamma}(n,n') = \int_{BZ'} d^2 \kv \, \Theta_{\alpha\beta\gamma}(\kv,n,n')$ and the $BZ'$ indicates that we are integrating in a small region where the quadratic approximation is enough to describe the bands of Graphene around the valleys with $\tau=\pm 1$. The limitations and proper choice of $BZ’$ is discussed in the main text.
In natural units $e = 0.0854$ and $694 \; \mathrm{eV}^2 = 1 \;\mathrm{T}$.
Finally $\lambda^\prime = 0.4 \; \mathring{A} \; \cdot v$ and $1\; \mathring{A} = \frac{1}{2000 \; eV}$, so that $\lambda = 0.06$.
Using this method for the calculation of the Green's function, the very big computational time required for calculations in magnetic fields can be overcome. I also approximated considering just few $k$-points per valley (since the $k$ points are weighted by a Gaussian function and they decay quite fast). Analyzing the structure of the matrix element $\Theta_{\alpha\beta\gamma}(n,n')$ it is easy to show that there is a selection rule on $\Delta n = n-n'$ for the allowed transitions, that is
$
\Delta n = \pm 1.
$
This becomes evident if we realize that the integral in the matrix element contains an integral over $k$ of Laguerre Functions, that are orthonormal by definition.
Expression \eqref{SHGmagneticfield0} can be rewritten in less compact, but more physically comprehensible way dividing the landau level number $n$ and the band index $s$ as follows

\begin{align}
&\Pi_{\alpha\beta\gamma}^{(2)}(\omega)=4 i \omega \sum_{\substack{n=1 \\ s=\pm 1}}^{+\infty} \Theta_{\alpha\beta\gamma}(n,-n-s)\left[n_{\mathrm{F}}(\varepsilon_n-\mu)-n_{\mathrm{F}}(-\varepsilon_{n+s}-\mu)\right] \left[ \frac{1}{(2i\omega)^2-\left(\varepsilon_n-\varepsilon_{n+s}\right)^2}+\frac{1}{(i\omega)^2-\left(\varepsilon_n-\varepsilon_{n+s}\right)^2}\right],
\label{SHGmagneticfield1}
\end{align}

The sum over $s$ accounts for the selection rule discussed in the previous paragraph and now the summation is taken just over a single positive Landau level index
.

\end{document}